\documentclass[aps,prl,noshowpacs,superscriptaddress,amsmath,amssymb,twocolumn,preprintnumbers,reprint]{revtex4}
\usepackage[utf8]{inputenc}
\usepackage{graphicx}
\usepackage{dcolumn}
\usepackage{bm}
\usepackage{amsmath}
\usepackage{amssymb}
\usepackage[english]{babel}
\usepackage{hyperref}
\usepackage{natbib}
\usepackage{xcolor}
\hypersetup{colorlinks=true,linkbordercolor=blue,linkcolor=blue, citecolor=blue,pdfborderstyle={/S/U/W 1}}
\usepackage{url}

\def\be{\begin{equation}}
 \def\ee{\end{equation}}
  \def\bea{\begin{eqnarray}}
 \def\eea{\end{eqnarray}}

\newcommand{\eq}[1]{(\ref{#1})}
\newcommand{\jpsi}{J/\psi}
\newcommand{\dd}{{\rm d}}
\newcommand{\sqrts}{\sqrt{s}}
\newcommand{\pt}{p_{_\perp}}
\newcommand{\dsigpp}{\dd\sigma_{\rm pp}}
\newcommand{\dsigAA}{\dd\sigma_{_{\rm AA}}}
\newcommand{\RAA}{\ensuremath{R_{_{\rm AA}}}}
\newcommand{\dydpt}{\dd{y}\,\dd\pt}
\newcommand{\dsigpphat}{\dd{\hat{\sigma}}_{\rm pp}}
\newcommand{\qhat}{\hat{q}}
\newcommand{\epsbar}{x}
\newcommand{\epsilonbar}{\bar{\epsilon}}
\newcommand\omc{\omega_c}
\newcommand\omcp{{{\bar{\omega}}_c}}
\newcommand\meanz{\langle z \rangle}
\newcommand\meaneps{\langle \epsilon \rangle}
\newcommand\meanepsbar{\langle \bar{\epsilon} \rangle}
\def\cO#1{{{\cal{O}}}\left(#1\right)}

\begin{document}
\title{Quenching of Hadron Spectra in Heavy Ion Collisions at the LHC}
\author{Fran\c{c}ois Arleo}
\affiliation{Laboratoire Leprince-Ringuet, \'Ecole polytechnique, CNRS/IN2P3,  Universit\'e Paris-Saclay,  91128, Palaiseau, France}

\date{\today}
\begin{abstract}
The $\pt$ dependence of the nuclear modification factor $\RAA$ measured in PbPb collisions at the LHC exhibits a universal shape, which can be very well reproduced in a simple energy loss model based on the BDMPS medium-induced gluon spectrum. The scaling is observed for various hadron species ($h^\pm$, $D$, $\jpsi$) in different centrality classes and at both colliding energies, $\sqrts=2.76$~TeV and $\sqrts=5.02$~TeV. Results indicate an 10--20\% increase of the transport coefficient from $\sqrts=2.76$~TeV to $\sqrts=5.02$~TeV, consistent with that of particle multiplicity. Based on this model, a data-driven procedure is suggested, which allows for the determination of the first and second moments of the quenching weight without any prior knowledge of the latter.
\end{abstract}
\maketitle

\setcounter{footnote}{0}
\renewcommand{\thefootnote}{\arabic{footnote}}

Since its discovery~\cite{Adcox:2001jp,Adler:2002xw}, jet quenching has almost become an iconic symbol for quark-gluon plasma (QGP) formation in high-energy heavy ion collisions. The depletion, with respect to pp collisions, of high transverse momentum particle production observed in AuAu and CuCu collisions at RHIC ($\sqrts=130$, $200$~GeV)~\cite{Arsene:2004fa,Back:2004je,Adams:2005dq,Adcox:2004mh} and more recently in PbPb collisions at the LHC ($\sqrts=2.76$, $5.02$~TeV)~\cite{Abelev:2012hxa,Aad:2015wga,CMS:2012aa,Khachatryan:2016odn} indeed signals the early formation of a dense medium in these collisions. Today, there is a general consensus in the community that the physical origin of jet quenching is the energy loss of quarks and gluons due to multiple scattering in QGP, as predicted long ago~\cite{Bjorken:1982tu,Wang:1991xy}. 

The wealth of measurements presently available, from single particle spectra (identified and all charged hadrons, heavy-quark mesons, quarkonia, jets) to double inclusive particle production ({\it e.g.}, jet-tagged or photon-tagged correlations) at RHIC and LHC has allowed for a sophisticated phenomenology on jet quenching; see Refs.~\cite{Majumder:2010qh,Mehtar-Tani:2013pia,Armesto:2015ioy,Qin:2015srf} for reviews. Most studies which aim at explaining data are based on a perturbative picture for medium-induced gluon radiation, however within a variety of theoretical frameworks and underlying assumptions. The description of the hot medium, often within (ideal or viscous) hydrodynamics, may also vary significantly from one study to another. Often these approaches are based on Monte Carlo event generators, which allow for the inclusion of different physical processes as well as for the computation of different observables within a consistent framework~\cite{Armesto:2009fj,Schenke:2009gb,Zapp:2012ak,Majumder:2013re}. This versatility is certainly appealing. However the complexity of these approaches could also make the interpretation of the experimental results delicate.

In this article, I would like to explore a somewhat different strategy by assuming only a single physical process -- radiative parton energy loss -- to describe the quenching of single inclusive hadron spectra at large transverse momentum ($10$~GeV $\lesssim \pt \lesssim 300$~GeV) within a simple analytic model. The goal is to extract, using the least number of assumptions,  the amount of energy lost by the partons inside the medium from the quenching of light hadrons, $D$ and $\jpsi$ mesons. The results are also compared to an independent data-driven approach, based on the Taylor expansion of the pp production cross section, which allows for computing the first moments of the energy loss probability distribution.
No attempt to model the produced medium is performed, here characterized by a single energy loss scale extracted from the high-precision measurements. In particular, I will not investigate the relationship between energy loss scales and medium properties, such as the local transport coefficient $\qhat({\bf x}, \tau)$. This has been performed in a series of studies which show a systematic deviation from the extracted transport coefficient and its perturbative estimate~\cite{Bass:2008rv,Armesto:2009zi,Renk:2011gj,Horowitz:2011gd,Burke:2013yra,Andres:2016iys}; other approaches nevertheless claim that RHIC or LHC data are consistent with perturbative expectations~\cite{Peshier:2008bg,Zakharov:2012fp}.

There are many reasons to focus on particle production at large $\pt$. First of all, cold nuclear matter effects (nuclear parton distributions, energy loss in cold nuclear matter or the Cronin effect due to multiple scattering) vanish as soon as the hard scale of the process exceeds the saturation scale(s) of the incoming nuclei, $Q=\cO{\pt} \gg Q_s^{\rm A}$, where $Q_s^{\rm A}\simeq 1$--$2$~GeV at the LHC (see the discussion in~\cite{Arleo:2016lvg}). In addition, various {\it hot} medium effects which could affect the production of particles at low $\pt$, say $\pt \lesssim 10$~GeV, are also expected to fade out when $\pt$ gets larger. This is the case of possible coalescence processes in QGP~\cite{Fries:2003vb} or the role of power-suppressed direct processes~\cite{Brodsky:2008qp,Arleo:2009ch}, to mention only a couple.
The quenching of particle production due to parton energy loss also becomes much simpler at large $\pt$.  In the high-energy limit, radiative energy loss becomes independent of the parton energy (up to logarithms)~\cite{Baier:1996kr,Gyulassy:1999zd,Gyulassy:2000fs,Wiedemann:2000za}. Perhaps more importantly, the power law behavior of the pp production cross section expected at large $\pt$ leads to a simple scaling property for $\RAA$ discussed below.
Finally, thanks to the considerable integrated luminosity acquired at the LHC, the measurement of $\RAA$ is now performed up to $\pt \simeq 300$~GeV~\cite{Khachatryan:2016odn,ATLAS-CONF-2017-012}, making possible the study of its $\pt$ dependence over two orders of magnitude. 

The usual way to model energy loss effects on the production of particle $i$ in heavy-ion collisions is to rescale its pp production cross section, $\dsigpp^{i}/\dd{y}\,\dd\pt$. The minimum bias (centrality integrated) heavy-ion production cross section thus reads~\cite{Baier:2001yt}
\be
\label{sig-AA} 
\frac{\dsigAA^{i}(\pt)}{\dydpt}  = A^2\,\int_{0}^{\infty}\,\dd\epsilon\  \frac{\dsigpp^{i}(\pt+\epsilon)}{\dydpt}\ {P}_{_{i}}(\epsilon, E)\ ,
\ee
where the so-called quenching weight, ${P}_{_{i}}$, represents the probability density for the particle $i$ with energy $E$ to lose the energy $\epsilon$ while traversing the hot medium.  It is characterized by the scale $\omc = 1/2\,\qhat\,L^2$, where $\qhat$ is the transport coefficient of the medium and $L$ its length. In the high energy limit, $\omc/E\to0$, medium-induced gluon radiation becomes independent of the parton energy, up to logarithmic corrections. As a consequence the quenching weight becomes a function of $\epsilon/\omc$ only, ${P}(\epsilon, E\gg \omc) \equiv 1/\omc\,\bar{P}(\epsbar\equiv\epsilon/\omc)$. Once $\omc$ is known, the mean energy loss $\meaneps=\langle x\rangle\,\omc$ can be determined, where $\langle x \rangle$ is the first moment of $\bar{P}$; $\langle x \rangle=1/2\,\alpha_s\,C_R$ for a propagating parton in color representation $R$~\cite{Baier:1998kq}.

In practice, however, the detected particle (typically a hadron) is not the particle that experiences the energy loss, a quark or a gluon. Introducing the fragmentation function of parton $k$ into the hadron $h$, the heavy-ion cross section can be written as (with $x\equiv z\,\epsilon/\omc$)
\bea
\label{sig-AA-2} 
\frac{\dsigAA^{h}(\pt)}{\dydpt}  &=& A^2\,\sum_k\,\int_0^1\,\dd{z}\,D_{k}^h(z)\int_{0}^{\infty}\,\dd\epsbar\,\nonumber\\ &&\frac{\dsigpphat^{k}(\pt/z+\omc\,\epsbar/z)}{\dydpt}\,\frac{1}{z}\,\bar{P}_{_{k}}(\epsbar/z)\,, \nonumber\\[-0.3cm] 
\eea
where $z$ is the momentum fraction carried away by the hadron. Eq.~(\ref{sig-AA-2}) involves the weighted sum over the flavour of the partons fragmenting into the detected hadron, in pp and AA collisions. In the massless scheme, the relative contributions to this incoherent sum depend essentially on the product of parton distribution functions, hidden in the cross section, $\dsigpphat^{k}/\dydpt$, and on the fragmentation functions. I shall assume here that only one partonic channel dominates the production of the measured hadrons.
This approximation should be appropriate at the LHC where the production of light hadrons is dominated by gluon fragmentation~\cite{Sassot:2010bh}, although at very large $\pt$, say $\pt \gtrsim 100$~GeV,
quark fragmentation processes are no longer be negligible. Regarding $D$ meson production, it is dominated by charm-quark fragmentation~\cite{Kniehl:2004fy}, while at large $\pt$ heavy-quarkonia are likely to be produced by the fragmentation of a gluon or color octet ${[Q\bar{Q}]}_8$ states~\cite{Ma:2014svb}.
%
\begin{figure}[t]
\begin{center}
    \includegraphics[width=8.3cm]{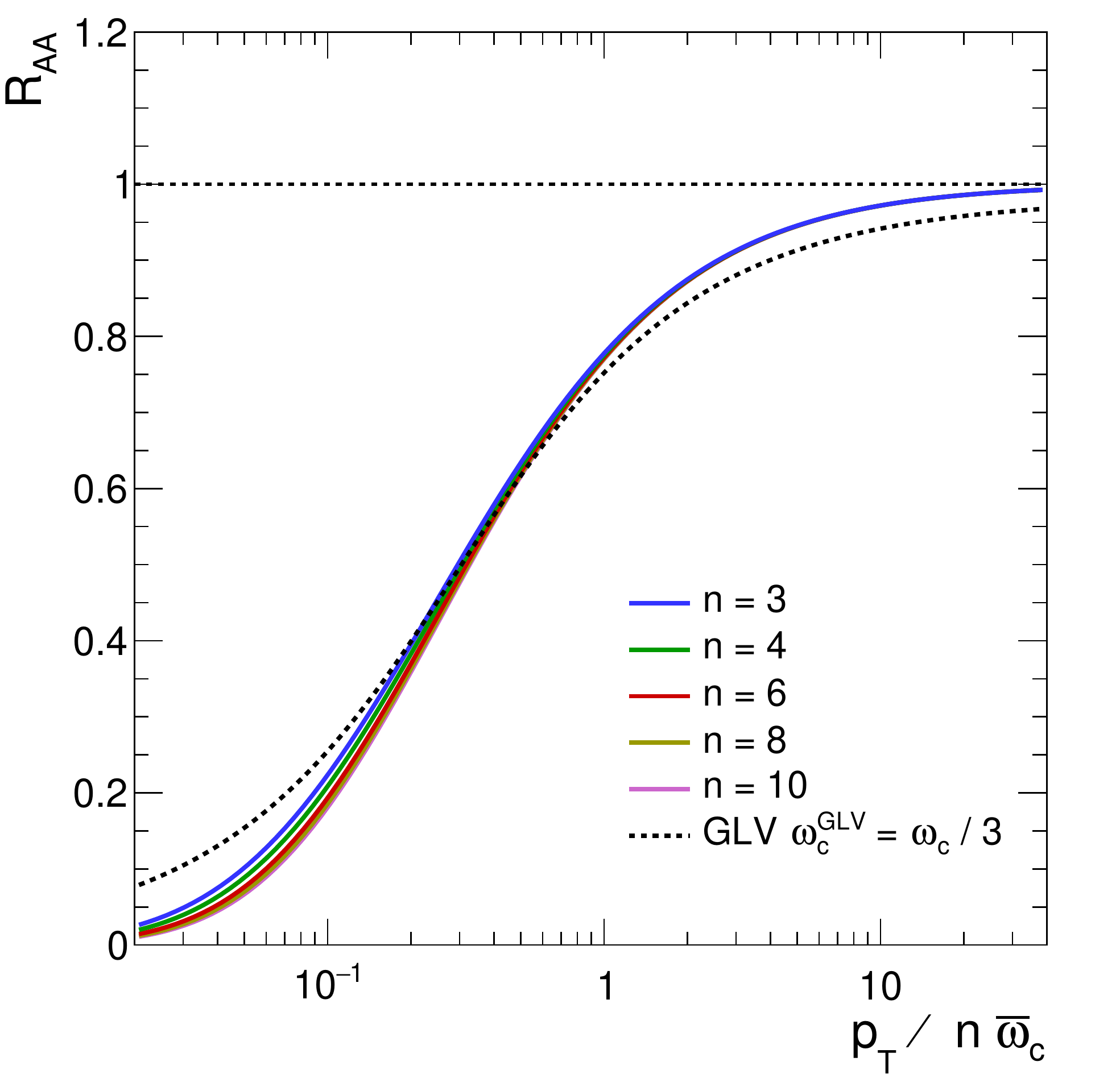}
  \end{center}
\vspace{-0.4cm}
\caption{Nuclear modification factor \RAA\ as a function of $\pt/\,n\omc$ for different values of the power law exponent.}
  \label{fig:exponents}
\end{figure}
Assuming $1/z\,\bar{P}(\epsbar/z)$ to be a smooth function of $z$ compared to the rest of the integrand, this term is evaluated at a typical $\meanz$ and taken out of the $z$ integral, leading to
\bea\label{sig-AA-approx} 
\frac{\dsigAA^{h}(\pt)}{\dydpt} &\simeq& A^2\ \int_0^{\infty}\,\dd\epsbar\  \frac{\dsigpp^{h}(\pt+\omcp\,\epsbar)}{\dydpt}\ \bar{P}(\epsbar)\,,\nonumber \\[-0.2cm]
&&
\eea
with $\omcp\equiv \meanz\,\omc$. From \eq{sig-AA-approx},  the nuclear modification factor,
\be\label{eq:RAA}
\RAA^{h}(\pt) \equiv \frac{1}{A^2}\ \frac{\dsigAA^{h}(\pt)}{\dydpt} \bigg/\, \frac{\dsigpp^{h}(\pt)}{\dydpt}\,,
\ee
can be determined. When $\pt$ is large with respect to any other scale in the process (say, $\Lambda_{_{\rm QCD}}$ or the heavy-quark mass $m_Q$), the single inclusive hadron spectrum in pp collisions exhibits a power law behavior, $\dsigpp^{h}/\dydpt\propto\pt^{-n}$ where $n$ depends in principle on the hadron species and on the center-of-mass energy of the collision, $n=n(h, \sqrts)$. According to this simple energy loss model, the nuclear modification factor at large $\pt$,
\be\label{eq:RAA2}
\RAA^{h}(\pt) = \int_0^{\infty}\,\dd\epsbar\  \left( 1 + x\,\frac{\omcp}{\pt} \right)^{-n}\ \bar{P}(\epsbar)\,,
\ee
 thus becomes a scaling function of 
 $\pt/\omcp$
 for a given power law index $n$. Such a scaling behavior can be observed from the centrality dependence (corresponding to different energy loss scales $\omcp$) of $\RAA^h(\pt)$ at a given energy $\sqrts$. 
 Performing the replacement $(1+x\,\omcp/\pt)^{-n}$ by $\exp(-x\,n\,\omcp/\pt)$, as suggested by Baier {\it et al.} (BDMS)~\cite{Baier:2001yt}, leads to an approximate scaling in the variable $\pt/\,n\omcp$, which now allows for comparing the values of $\RAA$ at different center-of-mass energies and/or for different hadron species. Note that within the BDMS approximation, the $\pt$ dependence of $\RAA$ is directly connected to the medium-induced gluon spectrum ($u=\omega/\omc$)~\cite{Baier:2001yt}
\be\label{eq:RAABDMS}
\RAA(y\equiv\pt/n\omcp) \simeq \exp\left[-\int_0^\infty\,\dd{u}\,\frac{d{I^\prime}(u)}{d{u}} \left(1-e^{-u/y}\right)\right]
\ee
with $d I^\prime(u)/d{u}=\omc\,d I(\omega)/d{\omega}$.

In the present article, $\RAA(\pt/\omcp)$ is computed numerically from \eq{sig-AA-approx} using the quenching weight computed in~\cite{Arleo:2002kh} from the BDMPS medium-induced gluon spectrum~\cite{Baier:1996kr,Baier:1996sk}. Fig.~\ref{fig:exponents} shows $\RAA$ as a function of $\pt/n\omcp$ for different values of power law exponents. As can be seen, scaling in $\pt/n\omcp$ is well observed, except at low $\pt/n\omcp$ and for the smallest values of $n$~\footnote{At LHC, however, the value of $n$ for the different particles species considered does not vary as much, $n\simeq 5$--$6$.}. It has also been checked, for consistency, that the BDMS analytic approximation, Eq.~\eq{eq:RAABDMS}, reproduces Eq.~\eq{eq:RAA2} well when $\pt / n\omc$ gets large. Finally, $\RAA$ is computed from \eq{eq:RAABDMS} using the GLV spectrum at first order in opacity~\cite{Gyulassy:1999zd,Gyulassy:2000fs}, 
shown as a dashed line in Fig.~\ref{fig:exponents}. For a meaningful comparison with BDMPS, the GLV energy loss scale has been rescaled by a factor 3, as already noted in~\cite{Salgado:2003gb}. Although the BDMPS and GLV medium-induced spectra behave somewhat differently in the infrared, respectively $u\,d I^\prime/du \propto 1/\sqrt{u}$ and $u\,d I^\prime/du \propto 1/\ln u$, the $\pt$ dependence of $\RAA$ is not too dissimilar; yet, the shape using the  GLV spectrum proves not as steep. 
\begin{figure}[tbp]
\begin{center}
    \includegraphics[width=8.3cm]{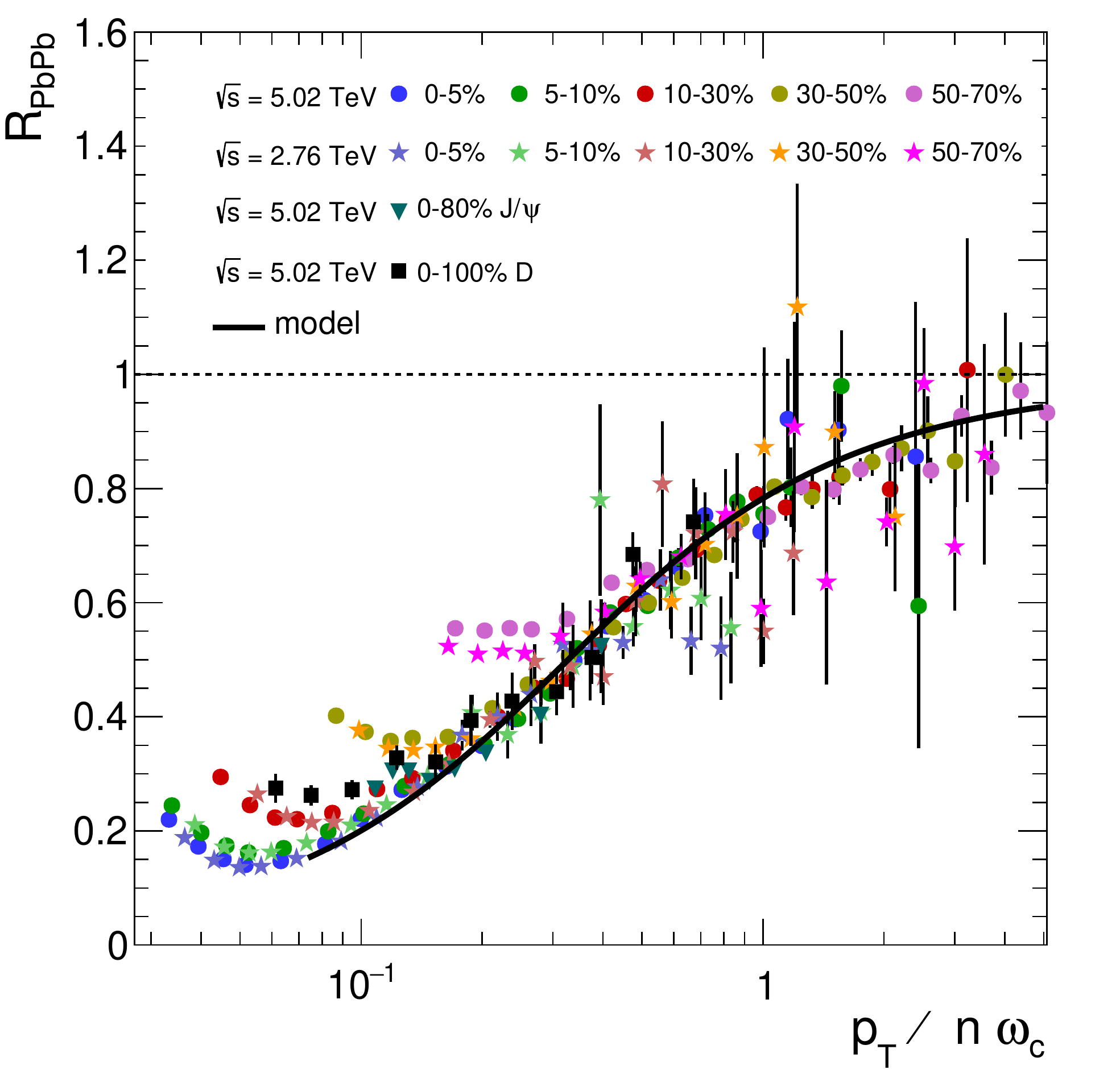}
  \end{center}
\vspace{-0.4cm}
\caption{$\RAA$ of $h^\pm$, $D$ and $\jpsi$ as a function of $\pt/\,n\omc$ in PbPb collisions at $\sqrts=2.76$~TeV and $\sqrts=5.02$~TeV in different centrality classes.}
   \label{fig:scaling}
\end{figure}

In this simple energy loss model, the {\it shape} of $\RAA$ as a function of $\pt$ is thus fully predicted once the exponent $n$ is known, obtained from a fit to the pp data at the corresponding center-of-mass energy. What remains to be determined is the energy loss scale $\omcp$, which is in principle a complicated (and virtually unknown)  function of the space-time evolution of the QGP energy density and the geometry of the heavy ion collision. Rather than modeling the hot medium, the value of $\omcp$ is obtained from `agnostic' 1-parameter fit to each data set, in a given centrality class and at a given $\sqrts$. Measurements include charged hadrons measured by CMS in five centrality classes~\footnote{No attempt is made to describe the data in the most peripheral class, $70$--$90\%$, which very different behavior remains unexplained.} at both colliding energies~\cite{CMS:2012aa,Khachatryan:2016odn}, $\jpsi$ and $D$ mesons measured respectively by ATLAS~\cite{ATLAS-CONF-2016-109} and CMS~\cite{CMS-PAS-HIN-16-001} at $\sqrts=5.02$~TeV in one centrality class, for a total number of 12 data sets. Data from ALICE~\cite{Abelev:2012hxa,Gronefeld:2016dtc,Adam:2015nna} are not included here as I focus on measurements with largest $\pt$, however these results will be included in the more detailed analysis~\cite{Arleo:2017ta}.

\begin{figure}[tbp]
\begin{center}
    \includegraphics[width=8.3cm]{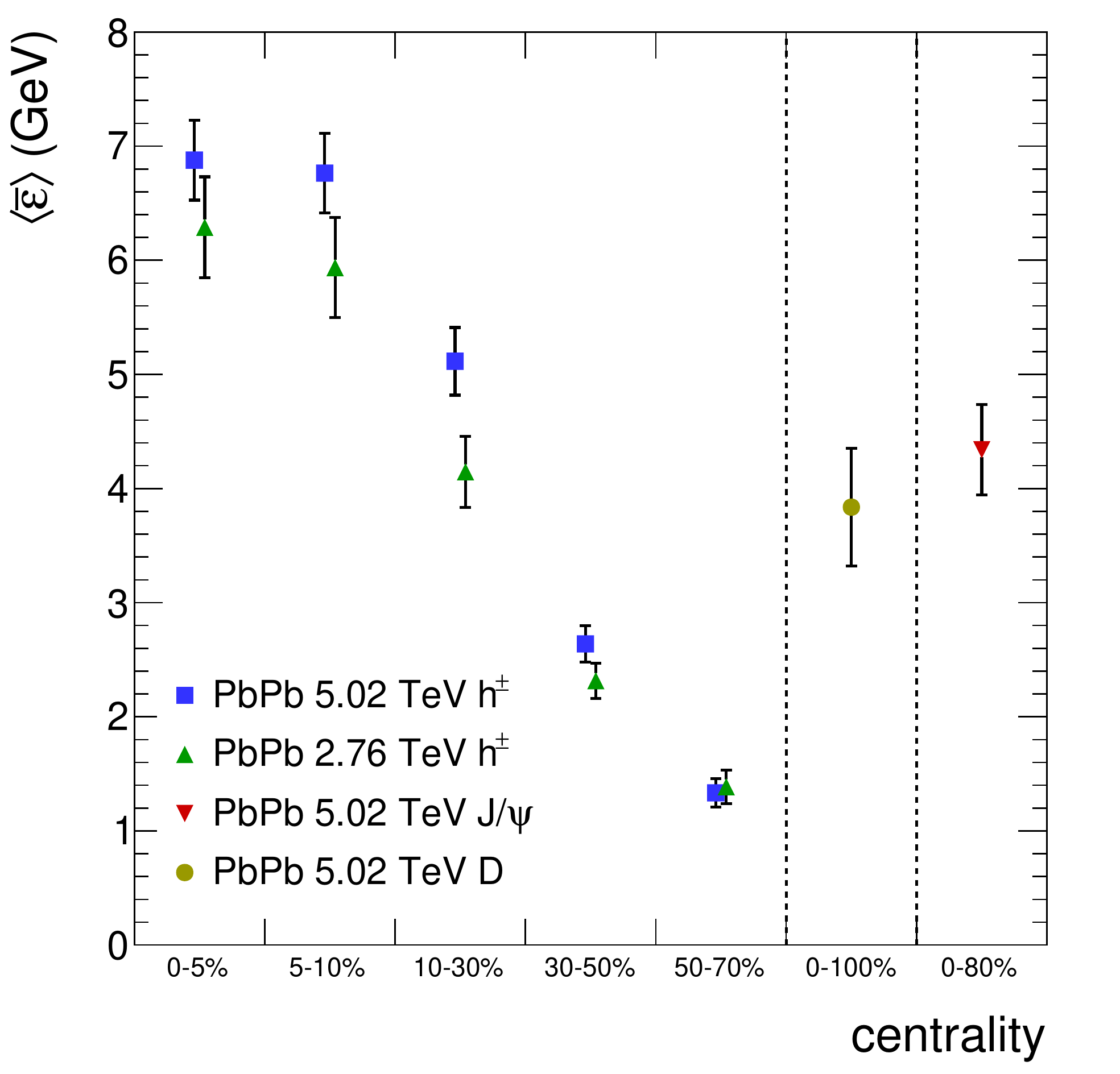}
  \end{center}
\vspace{-0.4cm}
\caption{Mean energy loss extracted in PbPb collisions at $\sqrts=2.76$~TeV (triangles) and $\sqrts=5.02$~TeV (squares) from the quenching of $h^\pm$, $D$, and $\jpsi$.}
  \label{fig:m1}
\end{figure}

The comparison of the fits to the individual data sets will be shown in a forthcoming publication~\cite{Arleo:2017ta}. Instead, Fig.~\ref{fig:scaling} shows all data points~\footnote{For clarity only statistical uncertainties are shown.} plotted as a function of the scaling variable, $\pt/\,n\omcp$, together with the shape of $\RAA$, Eq.~\eq{eq:RAA2}.
Clearly all data exhibit an almost perfect scaling, lining up into a single `universal' shape. This feature, predicted in the energy loss model and observed in data, supports the interpretation of a unique process responsible for the nuclear modification factors of all hadrons above a given $\pt$ in heavy-ion collisions at the LHC. In particular, I find it interesting that the quenching of heavy mesons ($D$ and $\jpsi$) obeys the exact same pattern, suggesting again that at large $\pt$ the same process affects similarly all hadron species, including bound states like heavy-quarkonia. Also worth to be noted are the scaling violations observed for lower $\pt$ particles. The lack of scaling emerges below $\pt \simeq 10$~GeV, for all centralities, which may signal the appearance of other phenomena below this energy scale.
Another remarkable fact is how well the shape observed in data coincides with that given by the energy loss model, based on the BDMPS medium-induced gluon spectrum. Note however that computing $\RAA$ using the GLV spectrum~\cite{Gyulassy:1999zd,Gyulassy:2000fs} also reproduces very well the shape of the data. The current data precision thus does not yet allow one or another gluon spectrum to be favored. Smaller systematic uncertainties at lower $\pt$ and improved statistics in the largest $\pt$ bins will help reach that goal.

\begin{figure}[t]
\begin{center}
    \includegraphics[width=8.3cm]{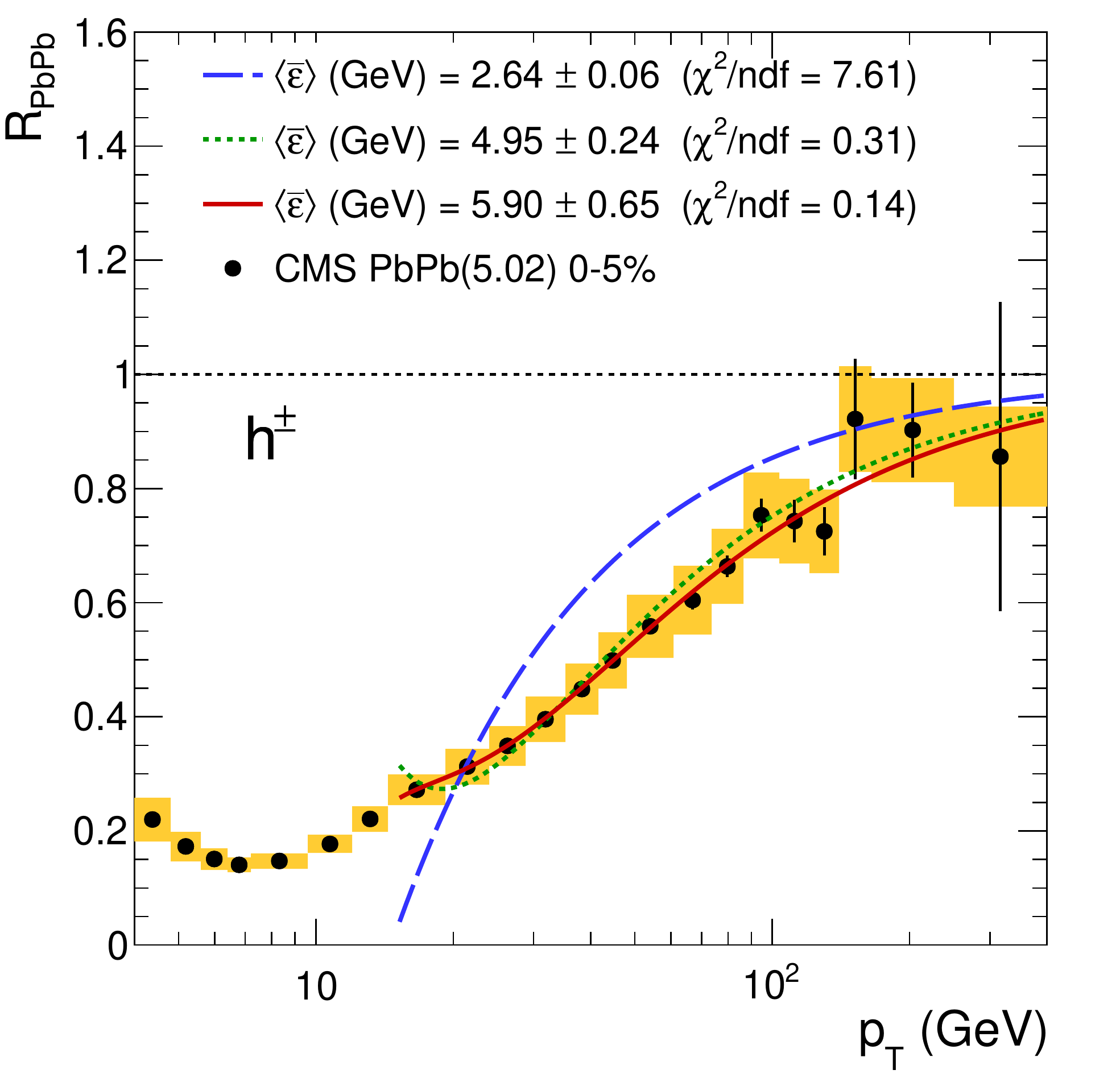}
  \end{center}
\vspace{-0.4cm}
\caption{Taylor expansion fits at first (dashed), second (dotted) and third order (solid) to the CMS measurements.}
  \label{fig:taylor}
\end{figure}

Apart from investigating the scaling of $\RAA$ for different hadron species or collision systems, this procedure allows for extracting the mean energy loss (times the mean fragmentation variable), $\meanepsbar \equiv \meanz\times\meaneps$, experienced by the fast parton in the QGP, as a function of centrality. Note that the error bars,  determined from the $\chi^2$ minimization procedure (with a usual tolerance criterion of $\Delta \chi^2=1$), are strongly correlated from one centrality class to another, since they arise mostly from the systematic uncertainties of the data. By changing the color prefactor of the induced spectrum from $C_F$ to $C_A$, I checked that the values of $\meanepsbar$ do not change significantly when the fragmenting parton is assumed to be a quark or a gluon.  The mean parton energy loss $\meaneps$ can be deduced from $\meanepsbar$ once $\meanz$ is known. It can be estimated from the (fractional) moments of the relevant fragmentation function, $\meanz \simeq \int \dd{z}\ z^{n+1}\ D_{k}^{h}(z) / \int \dd{z}\ z^{n}\ D_{k}^{h}(z)$, where $n$ is the power law exponent of the pp cross section.  At the LHC, NLO calculations indicate that $\meanz\simeq0.5$ for light hadron production~\cite{Sassot:2010bh}. 

As can be seen in Fig.~\ref{fig:m1}, $\langle \bar{\epsilon} \rangle$ is (as expected) maximal in the most central bins and of the order of 6--7~GeV. It starts to drop in the more peripheral classes, to reach $\langle \bar{\epsilon}\rangle \simeq1$~GeV in the $50$--$70\%$ centrality class. It is particularly interesting to note that the mean energy loss in PbPb collisions at $\sqrts=5.02$~TeV is roughly $10$--$20\%$ larger than at $\sqrts=2.76$~TeV. This is nicely consistent with the measurements of ALICE~\cite{Adam:2015ptt} which show that the multiplicity distribution, $dN/dy \propto \qhat \propto \meaneps$, increase by roughly the same amount.
The values of $\meanepsbar$ extracted from the quenching of $D$ and $\jpsi$ are also displayed in Fig.~\ref{fig:m1}, in similar centrality bins ($0$--$100\%$ and $0$--$80\%$, res\-pectively). Perhaps a bit surprisingly at first glance, no genuine difference between these two channels is observed. Assuming that $D$ and $\jpsi$ production at large $\pt$ come from charm quark and gluon fragmentation, respectively, a naive estimate  would  lead to ${\meaneps}_{\jpsi} / {\meaneps}_{D} = C_A/C_F=9/4$. The typical values of $\meanz$ for both particles  could of course be different, although not significantly since the $c\to{D}$ and $g\to\jpsi$ (or ${[c\bar{c}]}_8\to\jpsi$) fragmentation functions are hard~\cite{Mele:1990cw,Braaten:1993rw,Ma:2013yla}. However, as noted in~\cite{Kniehl:2004fy}, the gluon fragmentation component into $D$ mesons is not negligible, of the order of $30\%$ at the Tevatron. Also, the presence of color singlet states ${[c\bar{c}]}_1$ fragmenting into $\jpsi$, unaffected by the medium, could reduce the value of $\meaneps$ extracted from $\jpsi$ $\RAA$. This might explain, at least partly, why the estimates from $D$ and $\jpsi$ are not too dissimilar.

\begin{figure}[t]
\begin{center}
    \includegraphics[width=8.3cm]{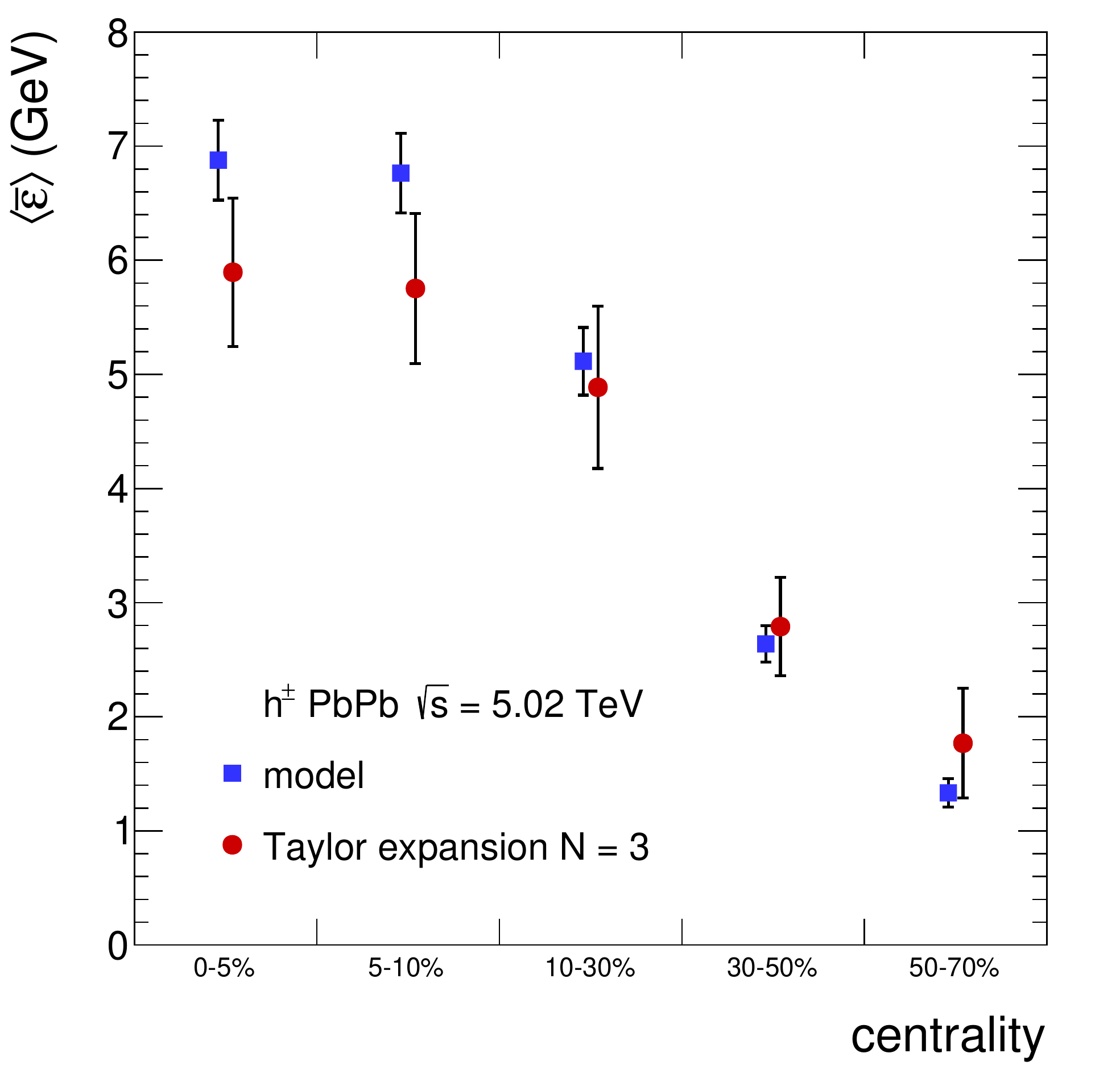}
  \end{center}
\vspace{-0.4cm}
\caption{$\meanepsbar$ in PbPb collisions at $\sqrts=5.02$~TeV from the BDMPS model (squares) and the Taylor expansion procedure (circles).}
  \label{fig:m1taylor}
\end{figure}

The above estimates of the mean energy loss rely on a model using a specific quenching weight. The success of a simple model, which predicts rightfully the shape of $\RAA(\pt)$ observed in data, gives some confidence on the robustness of these estimates. It is nevertheless legitimate to question the model dependence of these results. In the following, I suggest extracting the first moments of the quenching weight, without any prior knowledge of the latter. 

At the LHC, measurements of $\RAA$ extend for the first time at transverse momenta significantly larger than the typical energy loss scale, $\pt\gg\meaneps\simeq5$--$10$~GeV even in the most central collisions (see Fig.~\ref{fig:m1}). From the Taylor expansion of \eq{eq:RAA2},  $\RAA$ is conveniently approximated at large $\pt$ by
\be\label{eq:TaylorRAA}
\RAA^{^{(N)}}(\pt,\{\langle\epsilonbar^j\rangle\}) = \sum_{i=0}^N\,(-1)^i\,\frac{\Gamma(n+i)}{\Gamma(i + 1)\,\Gamma(n)}\, \frac{\langle\epsilonbar^i\rangle}{\pt^i}\,.
\ee
The first $N$ moments $\langle\epsilonbar^j\rangle$ can therefore be determined from a fit to $\RAA$ data, without any specific assumption on the quenching weight. In order to check this procedure, $\RAA$ toy data were generated according to a known quenching weight (in this case, that based on BDMPS gluon spectrum) and then fitted using \eq{eq:TaylorRAA} at orders $N=1,2,3$. This study shows that the first and second moments of the known quenching weight can be retrieved with a good accuracy from $\RAA^{^{(N=3)}}$~\cite{Arleo:2017ta}. Based on this result, the mean energy loss $\langle\epsilonbar\rangle$ has been determined for each LHC data sets discussed here. Fig.~\ref{fig:taylor} shows the result of the fits to the charged hadron $\RAA$ measured in the $0$--$5\%$ most central PbPb collisions. As can be seen, the quality of the fits improves as $N$ is getting larger. These totally `agnostic' estimates of the mean energy loss are compared in Fig.~\ref{fig:m1taylor} to the results based on the BDMPS model. Both approaches yield nicely consistent values, yet the uncertainties using the Taylor series tend to be larger because of the higher number of parameters used. 

In summary, I discussed the quenching of single hadron spectra in heavy-ion collisions at the LHC within a simple energy loss model. It is shown that the $\RAA$ of charged hadron, charmed mesons and heavy-quarkonia, measured at different $\sqrts$ and in various centrality classes scale on a single curve which is nicely predicted by the model. This allows for the extraction of the mean parton energy loss. Finally I also suggest a model-independent procedure to extract this quantity (and its variance), which gives consistent results.

\begin{acknowledgements}
I thank Matt Nguyen, St\'ephane Peign\'e and Marta Verweij for useful comments on the manuscript.
\end{acknowledgements}

\end{document}